\documentstyle[amsmath,amssymb,graphicx]{article}

\def\be{\begin{eqnarray}}
\def\ee{\end{eqnarray}}

\hoffset= - 1.0in         

\begin{document}

\hfill ITEP/TH-15/10

\bigskip

\centerline{\Large{
Black Hole Motion in Entropic Reformulation of General Relativity
}}

\bigskip

\centerline{A.Morozov}

\bigskip

\centerline{{\it ITEP, Moscow, Russia}}

\bigskip

\centerline{ABSTRACT}

\bigskip

{\footnotesize
We consider a system of black holes
-- a simplest substitute of a system of point particles
in the mechanics of general relativity --
and try to describe their motion with the help of
entropic action: a sum of the areas of black hole
horizons.
We demonstrate that such description is indeed consistent
with the Newton's laws of motion and gravity,
modulo numerical coefficients, which coincide but seem
different from unity.
Since a large part of the modern discussion of
entropic reformulation of general relativity is actually
based on dimensional considerations,
for making a next step it is crucially important to
modify the argument, so that these dimensionless
parameters acquire correct values.
}

\bigskip

\bigskip

A recent paper \cite{Ver} by Eric Verlinde
attracted a new attention to the old idea of
entropic reformulation of general relativity
\cite{Bek}.
Naturally, this paper provoked an avalanche
of new publications, we list just a few in
\cite{Verfirst}.
In this letter we provide still another
small illustration
of how this idea could work,
and what kind of discrepancies could
arise in attempts to formulate and validate
it at quantitative level.

\bigskip

{\bf 1. The problem and simplifying assumptions.}

In ordinary approach to quantum gravity one integrates
over gravitational fields $\{g\}$ with the weight defined
by Einstein-Hilbert (or Palatini) action $S\{g\}$
and some measure $d\mu\{g\}$:
\be
Z = \int e^{S\{g\}/\hbar}d\mu\{g\}
\ee
Different formulations of quantum gravity, from superstring theory
to loop gravity, make use of
different realizations (and, perhaps, generalizations)
of $S\{g\}$.
At the same time, quantum gravity is sometime believed to be
a {\it topological} theory, and topological theories are those
which do not have a non-trivial action, $S=0$,
only the measure $d\mu$.
This motivates the search for a {\it reformulation} of general
relativity, where Einstein action will be substituted by
some measure $d\mu$, which hopefully will be of
pure geometric nature.
From the point of view of thermodynamics,
$Z = e^{F/T}$, and $dF = TdS-dE$, where the first term,
$TdS$, is associated with the measure $d\mu$, while
the second, $dE$,-- with the action $S$.
In topological theories $dE=0$ and $Z = e^S$
is defined by pure {\it entropic} considerations.
Thus, if gravity is believed to be a topological theory,
it is actually believed to be pure entropic:
instead of Einstein action one can use just the entropy function.

Usually the simplest system to formulate and explore the dynamical
principles of the theory is a collection of point particles.
In the case of gravity there are no point particles:
the simplest objects which exist in this theory are black holes.
Thus the simplest toy model in gravity is a collection
of black holes. The entropy function for this system
is defined by the sum of areas of the black hole horizons,
i.e. {\it entropic action} is simply
\be
S = \varkappa\hbar \cdot L_{Pl}^{2-D}\sum_i A_i
\label{entroac}
\ee
where $A_i$
is the area of the horizon of the $i$-th black hole,
$M_{Pl} = L^{-1}_{Pl}$ is the Planck constant and the
measure $d\mu$ is now trivial.
We write the action in $D$ space-times dimensions,
to have one extra parameter, which can enter expressions
for dimensionless coefficients in what follows.

Of course, in entropic formulation there is no space-time,
thus $D$ is no more than a free parameter of the theory.
Moreover, there is also no Plank constant $\hbar$ in $e^{S/\hbar}$,
all dynamics in topological theory is pure {\it combinatorial}:
dictated by counting of degrees of freedom (basically
a calculus of integers, a section in {\it number theory},
perhaps, subjected to one or another regularization when
$D$ exceeds $2$).
The minimal action principle is still applicable in
the classical approximation, when  $S$ is large, i.e.
when all the distances, i.e. Schwarzschild radia and
distances between the black holes are large as compared
to the Planck length $L_{Pl}$.
In what follows we keep $L_{Pl}=1$ to simplify the formulas.

To promote the geometric formula (\ref{entroac})
into a real dynamical principle
one needs to specify, how the areas $A_i$ depend on the
state of the system, i.e. on location of our black holes
and their velocities.
Different versions of gravity theories (say, different
modifications of Einstein-Hilbert or Palatini actions)
can provide different expressions, but all of them are
quite sophisticated and non-local.
They are, however, drastically simplified for remote
black holes, when distances between them are much bigger
then their Schwarzschild radia, and this is the case
that we are going to analyze in the present letter.

Moreover, we make two further simplifying assumptions.
First, we define the horizons by Laplace's principle:
that the second cosmic (parabolic) velocity is equal to the
light speed $c=1$. Second, most disputable,
we assume that the shape of the horizon of a moving black hole
is deformed by Lorentz contraction in the longitudinal direction
(e.g. a spherical horizon for an isolated black hole becomes
an axially symmetric ellipsoid).
These assumptions make calculations trivial and provide
a clear picture of what happens without going deep
into sophisticated analysis of non-linear gravity.

\bigskip

{\bf 2. Kinematics: the second Newton's law.}

In general relativity our particles (black holes)
are never at rest: there is no parameter (like large mass)
which could be adjusted to keep them in a given position.
Moreover, they are necessarily accelerated.
Thus the first question to ask is where acceleration is
in the action principle (\ref{entroac}).
The answer is already suggested in \cite{Ver}:
for every particular {\it probe} black hole we have
a kind of a Newton's second law,
\be
m\vec a = -\varkappa\, T\,\frac{\delta A}{\delta \vec x}
\label{Nl}
\ee
where $m$ is the mass, $T$ the temperature
(inverse of Schwarzschild radius $r$), $A$ the area
$A\sim r^{D-3}$, $\varkappa A$ the entropy,
and $x$, $v$ and $a$ are position, velocity
and acceleration of the black hole.

We, however, prefer to interpret/derive the "Newton law"
(\ref{Nl}) in a somewhat different way from \cite{Ver}:
just as a simple kinematical relation.
Namely, imagine that our particle (black hole) just
started to move. Then during the time $\delta t$ it passes
the distance
\be
\delta x = \frac{a\delta t^2}{2}
\ee
then
\be
a\delta x = \frac{v^2}{2}
\ee
is expressed through velocity $v = a\delta t$ that it
finally achieved.
At the same time, the horizon of the black hole is now
deformed by the Lorentz contraction:
instead of a sphere with the area $A=\pi r^2$ it is
now an ellipsoid with the smaller area
\be
A+\delta A = A(1-C_{LC}v^2)
\ee
Thus
\be
a\delta x = \frac{v^2}{2}  = -\frac{1}{2C_{LC}}\cdot\frac{\delta A}{A}
= -C_{NL}\frac{\varkappa\, T\delta A}{m}
\label{Nl'}
\ee
where
\be
\boxed{
C_{NL} = \frac{m}{\varkappa AT}\cdot\frac{1}{2C_{LC}}
}
\label{cnl}
\ee
If black hole was already moving then $v$ is not infinitesimally
small, and $\delta \vec x = \vec v\delta t$.
Still $\vec a\delta \vec x = \vec a\vec v\delta t=\vec v\delta \vec v$, while
$\delta A = -2C_{LC}A\vec v\delta \vec v$, so that (\ref{Nl'})
is preserved with the same value of $C_{NL}$.
This is consistent with (\ref{Nl}) -- though is a somewhat weaker,
a scalar rather then a vector relation --
up to numerical constant $C_{NL}$, which still needs to be evaluated.

Evaluation of $C_{NL}$ seems quite important.
The point is that (\ref{Nl}) can actually be written on
pure dimension grounds -- provided one wants to find {\it some}
relation of this kind.
What could take us further, beyond pure dimensional consideration,
and thus provide a real {\it quantitative} argument in support
of entropic-reformulation ideas,
is evaluation of dimensionless numerical coefficients.
This is what makes any practical way to calculate $C_{NL}$
so interesting. Of course, in the case of our simple model this
is straightforward: to find $C_{NL}$
we need to know the Lorentz-contraction
factor $C_{LC}$ and the ratio $\varkappa AT/m$.
This will be the subject of the sections 4 and 5 below.

Note that there was no reference to the action principle (\ref{entroac})
in this section: eq.(\ref{Nl'}) is a pure kinematical relation.

\bigskip

{\bf 3. Dynamics: the Newton's gravity law.}

The action plays role when we study the interaction of several
black holes. Then there are two competing effects. First,
black hole horizon is deformed in the presence of gravitational field of
the other black holes, this leads to increase of the area.
Second, the field accelerates the black hole, what decreases
the area due to Lorentz contraction.
The minimal action principle requires that the two effects
exactly compensate each other.

Laplace principle easily defines the horizon of two black holes at rest:
\be
\frac{m_1}{|\vec r-\vec x_1|^{D-3}} +
\frac{m_2}{|\vec r-\vec x_2|^{D-3}} = C_{pot}^{-1}
\label{hor2}
\ee
$C_{pot}$ is a $D$-dependent Newton's constant, whose exact value
is irrelevant in consideration of this section.

If the distance $R$ between the black holes is much bigger
than their Schwarzschild radia then in the first approximation we get
\be
\frac{m_1}{r_1^{D-3}} = C_{pot}^{-1} - \frac{m_2}{R^{D-3}}
\label{rR}
\ee
i.e.
\be
A_1 = \Omega_{D-2}r_1^{D-2} =
\Omega_{D-2}\left(\frac{m_1}{C_{pot}^{-1}-{m_2}/{R^{D-3}}}
\right)^{\frac{D-2}{D-3}}
\ee
The closer the black holes the bigger are their horizons.

Under a small shift $\delta \vec x_1$ of the black hole
in space its horizon area changes:
\be
\delta A_1 \left( (D-2)\frac{\delta r_1}{r_1}
- 2C_{LC}(\vec a_1 \delta \vec x_1)\right) = 0
\label{varac}
\ee
The second item is the effect of Lorentz contraction and
the sum of two terms vanishes on equation of motion for
(\ref{entroac}). We assume here that the equations of motion
for different black holes are fully separated.

From (\ref{rR}) we have:
\be
\frac{m_1}{r_1^{D-3}}\frac{\delta r_1}{r_1} = -\frac{m_2}{R^{D-2}}\delta R
\ee
where $R\delta R = \vec R\delta \vec x_1$. Thus (\ref{varac}) is consistent
(again up to a difference between scalar and vector equations) with
the Newton's gravity law:
\be
\vec a_1 = -C_{GL}\vec\nabla_1\left(\frac{C_{pot} m_2}{R^{D-3}}\right)
\ee
with
\be
\boxed{
C_{GL} = \frac{D-2}{D-3}\cdot\frac{1}{2C_{LC}}
}
\label{cGL}
\ee

\bigskip

{\bf 4. Black hole numerology.}

While $C_{GL}$ in (\ref{cGL}) depends on nothing but the Lorentz contraction
factor $c_{CL}$, the kinematical factor $C_{NL}$ in (\ref{cnl})
is different: it involves detailed information about the black hole physics
\cite{Zel}-\cite{Akh}.

As explained at the end of section 1, we define the Schwarzschild radius $r$
by equating the parabolic velocity to $c=1$, i.e. from the condition
\be
C_{pot}\frac{m}{r^{D-3}} = 1
\label{sr}
\ee
We already used a more involved version of this equation in (\ref{hor2}).

Parameter $C_{pot}$ is normalization of a Green function for Laplace equation
in $D-1$ dimensions,
$\Delta_{D-1} \left(C_{pot}\frac{m}{r^{D-3}}\right) = C_L m\delta^{(D-1)}(\vec r)$.
The Gauss law then implies, that
\be
C_{pot}(D-3)\Omega_{D-2} = C_L
\ee

The Hawking temperature is
\be
T = \frac{C_T}{r}
\ee
where $C_T$ can be defined from quasiclassical considerations \cite{HawT},
a simple version of such derivation is recently analyzed in \cite{Akh}.

The area of the spherical horizon in the rest frame of the black hole is
\be
A = \Omega_{D-2}r^{D-2}
\ee
where the angular integral
\be
\Omega_{D-2} = \frac{2\pi^{\frac{D-1}{2}}}{\Gamma\left(\frac{D-1}{2}\right)}
\ee

Finally, the entropy of the black hole is proportional, which presumably
enters the r.h.s. of (\ref{Nl}), is proportional but not equal to the area,
\be
{\rm Entropy} = S/\hbar = \varkappa A
\ee

In terms of these parameters the ratio
\be
{\kappa AT}{m} = \varkappa C_T C_{pot}\Omega_{D-2} = \frac{\varkappa C_TC_L}{D-3}
\ee
The values of the parameters for $D=4$ are known since \cite{HawT}:
\be
D=4: \ \ \ \ \ C_{pot} = 2, \ \ \ \ \varkappa = \frac{1}{4},\ \ \ \ C_T = \frac{1}{4\pi},
\ \ \ \ \Omega_2 = 4\pi \ \ \ \ \Longrightarrow\ \ \ \ \frac{\varkappa AT}{m} = \frac{1}{2}
\ee
Generalization to arbitrary $D$ is now available in many papers, see, for example,
\cite{hD}. We borrow concrete formulas from a recent review \cite{Lan}:
\be
C_{pot}\Omega_{D-2} = \frac{16\pi}{D-2},
\ \ \ \ \varkappa = \frac{1}{4},
\ \ \ \ \ C_T = \frac{D-3}{4\pi}
\ \ \ \ \Longrightarrow\ \ \ \ \frac{\varkappa AT}{m}
= 4\varkappa\cdot\frac{D-3}{D-2}
= \frac{D-3}{D-2}
\ee
Substituting this ratio into (\ref{cnl}), we obtain:
\be
\boxed{
C_{NL} = \frac{D-2}{D-3}\cdot\frac{1}{2C_{LC}} \
\stackrel{(\ref{cGL})}{=}\ C_{GL}
}
\label{cc}
\ee
Remarkably, the two coefficients $C_{NL}$ and $C_{GL}$,
which both need to be unities for entropic principle to work,
at least coincide for arbitrary space-time dimension $D$.

\bigskip

{\bf 5. The Lorentz-contraction factor and parameters $C_{NL}$ and $C_{GL}$.}

To check if the common value of the two parameters is unity or not,
we need to evaluate the Lorentz-contraction factor $C_{LC}$.
It defines the deviation relative area
of the surface of the axially symmetric ellipsoid
$x_1^2+\ldots +x_{D-2}^2 + \gamma^2z^2=1$
with $\gamma^{-1} = \sqrt{1-v^2}$ and $\Omega_{D-2}$ from unity for $v^2\ll 1$:
\be
2C_{LC} = \frac{\int_0^\pi \sin^{D-1}\theta d\theta}
{\int_0^\pi \sin^{D-3}\theta d\theta} = \frac{D-2}{D-1}
\label{cLC}
\ee
Substituting this value into (\ref{cc}) we finally obtain:
\be
\boxed{\
C_{NL} = C_{GL} = \frac{D-2}{D-3}\cdot\frac{1}{2C_{LC}}
= \frac{D-1}{D-3}\ \neq 1\
}
\label{cc1}
\ee

\bigskip

{\bf 6. Conclusion.}

The main goal of this letter was explicit evaluation
of two numerical coefficients: $C_{NL}$ and $C_{GL}$
in the second (\ref{Nl'}) and gravitation (\ref{cGL}) laws
respectively.
For entropic reformulation of general relativity to work
in its most naive form, based on the action principle
(\ref{entroac}), these two coefficients should be equal to
unity. We did not manage to adjust them in this way,
moreover, our answers depend non-trivially on the free
parameter $D$ -- the space-time dimension.
Remarkably, though not unities, the two coefficients are
the same, and both discrepancies can be simultaneously cured
if one slightly changes the definition of the Lorentz-contraction
factor, from (\ref{cLC}) to $\frac{D-2}{D-3}$,
for example, by {\it postulating} the velocity-dependence
of the action (\ref{entroac}) in the form
$A_{{\rm rest}}\left(1 - \frac{D-2}{D-3}\cdot\frac{v^2}{2} + O(v^4)\right)$.
Such {\it ad hoc} postulates would, however,
decrease the attractiveness of the entire approach
and therefore are undesirable.
Before one can move further with {\it quantitative} development
of entropic reformulation along the lines of our section 1,
which would include gravitational
radiation and corrections beyond classical
(small $L_{Pl}$) approximation, and their comparison with
various programs of gravity quantization,
it is necessary to find and correct the mistakes
(arithmetical or conceptual)
in the simple calculations, described in above sections 2-5.
The next small step would be to consider the next
corrections in $v/c$ and $r/R$, also including non-linear
effects of general relativity (like those, responsible for
perihelion shift and the Lamb shift of orbital frequencies \cite{LBO}).

\section*{Acknowledgements}

I am indebted for stimulating discussions of related subjects to
E.Akhmedov, S.Apenko and A.Mironov.
I also appreciate corrections and advice by P.Burda, D.Diakonov,
V.Shevchenko and A.Sleptsov.
Of course, they are in no way responsible for possible mistakes in the
present paper.

The work was partly supported by Russian Federal Nuclear Energy Agency,
by Russian Education and Science Ministry under the contracts 02.740.5029
and 02.740.11.5194,
by RFBR grant 10-02-00499, by joint grants RFBR-CNRS 09-01-93106,
09-01-92440-CE, 09-02-91005-AFN,
09-02-90493-Ukr, 10-02-92109-Yaf-a,
and by the Russian president's grant for support of the scientific schools
NSh-65290.2010.2.

\end{document}